# Phase-Modulated Elastic Properties of Two-Dimensional Magnetic FeTe: Hexagonal and Tetragonal Polymorphs

Yunfei Yu,[1] Mo Cheng,[2] Zicheng Tao,[3,4] Wuxiao Han,[1] Guoshuai Du,[1] Yanfeng Guo,[3,4] Jianping Shi,[2] Yabin Chen[1,5,6,*]

[1]*School of Aerospace Engineering, Beijing Institute of Technology, Beijing 100081, P.R. China*

[2]*The Institute for Advanced Studies, Wuhan University, Wuhan 430072, P.R. China*

[3]*School of Physical Science and Technology, ShanghaiTech University, Shanghai 200031, P.R. China*

[4]*ShanghaiTech Laboratory for Topological Physics, Shanghai 201210, P.R. China*

[5]*Advanced Research Institute of Multidisciplinary Sciences, Beijing Institute of Technology, Beijing 100081, P.R. China*

[6]*BIT Chongqing Institute of Microelectronics and Microsystems, Chongqing, 400030, P.R. China*

[*]Corresponding author: Y.C. (chyb0422@bit.edu.cn)

**ABSTRACT:** Two-dimensional (2D) layered magnets, such as iron chalcogenides, have emerged these years as a new family of unconventional superconductor and provided the key insights to understand the phonon-electron interaction and pairing mechanism. Their mechanical properties are of strategic importance for the potential applications in spintronics and optoelectronics. However, there is still lack of efficient approach to tune the elastic modulus despite the extensive studies. Herein, we report the modulated elastic modulus of 2D magnetic FeTe and its thickness-dependence via phase engineering. The grown 2D FeTe by chemical vapor deposition can present various polymorphs, i.e. tetragonal FeTe ($t$-FeTe, antiferromagnetic) and hexagonal FeTe ($h$-FeTe, ferromagnetic). The measured Young's modulus of $t$-FeTe by nanoindentation method showed an obvious thickness-dependence, from 290.9±9.2 to 113.0±8.7 GPa when the thicknesses increased from 13.2 to 42.5 nm, respectively. In comparison, the elastic modulus of $h$-FeTe remains unchanged. Our results could shed light on the efficient modulation of mechanical properties of 2D magnetic materials and pave the avenues for their practical applications in nanodevices.

# INTRODUCTION

Two-dimensional (2D) layered nanomaterials have attracted extensive attention nowadays and exhibited the tremendous potential for applications in nanotransistors [1-3], nanomechanics [4, 5], biomedical devices [6-12], and flexible imaging [13-16], owing to their distinguished elastic properties and planar structures [17]. For instance, it is found that the Young's modulus and ultimate strain of monolayer graphene reached up to 1.0 TPa and 6.4%, respectively, much better than those of the conventional bulk semiconductors and alloys [18]. Interestingly, black phosphorus, as a representative anisotropic 2D model, presents the lattice orientation-dependent mechanical properties because of the in-plane puckered structure, that is, the elastic modulus is remarkably larger along zigzag (58.6 GPa) than armchair (27.2 GPa) directions [19]. However, it is noted that the efficient strategy to modulate the intrinsic mechanical properties of 2D materials has remained exclusive, despite much effort devoted over the past decade. In this regard, point defect like vacancy or dopant, nanobubbles and linear wrinkles have been intentionally explored to tailor the mechanical stiffness and tensile strain of $MoS_2$ and graphene, while those structural defects can play the essential role of scattering centers and hence inevitably affect their carrier mobility and electrical performances [20].

Polymorph, as a novel degree of freedom, can be exploited to modulate the physical properties of a given structural material without changing its stoichiometry. Meanwhile, it is well known that many 2D layered nanomaterials can present various polymorphic structures due to their different stacking orders. For example, trilayer graphene with 3R stacking geometry breaks both the inversion symmetry and isospin symmetry, leading to the remarkable strongly correlated behavior and further tunable superconductivity, which was found only in twisted bilayer graphene with a rigorous magic angle [21]. The in-plane mechanical properties of 2D materials are speculated to be closely related to their atomic configurations. $MoTe_2$, as a typical transition metal dichalcogenide, can exist thermodynamically in 2H (isotropic and hexagonal symmetry), 1T′ (anisotropic and distorted octahedral structure), and $T_d$ (anisotropic and octahedral structure) phases. Interestingly, it is reported that the elastic modulus of these phases are $110\pm16$, $99\pm15$, and $102\pm16$ GPa, respectively [22]. Therefore, the identical chemical composition offers the diverse mechanical properties due to the distinct polymorphs and symmetries.

As a novel paradigm of quantum materials, it is well known that 2D iron-based chalcogenides (FeS, FeSe, and FeTe) offer a distinct landscape to investigate their anomalous metallic states and even unconventional superconductivity [23]. Taking FeTe as an example, the phase-tunable growth approach via chemical vapor deposition was achieved to prepare the layered tetragonal FeTe (*t*-FeTe) and non-layered hexagonal FeTe (*h*-FeTe) with controlled thickness [24, 25]. The different phases result in the antiferromagnetic and ferromagnetic properties of *t*-FeTe and *h*-FeTe, respectively, and the Curie temperature of *h*-FeTe reaches up to ~220 K when its thickness is 30 nm. Moreover, iron-based chalcogenides stand in the special position due to their novel quantum phenomena, and especially there is still lack of reports on the superconductivity of FeTe [26, 27]. To our knowledge, the in-plane Young's modulus of *t*-FeTe and *h*-FeTe and their thickness-dependence remain unrevealed. We believe that the measurement and modulation of mechanical properties of FeTe can help to understand the transport dynamics of carriers and the potential superconducting behavior.

To date, the nanoindentation method based on atomic force microscope (AFM) has been widely utilized to characterize the mechanical properties of 2D layered nanomaterials. In this work, we report the modulated elastic properties of 2D magnetic FeTe via phase engineering. The extracted modulus of *t*-FeTe by AFM nanoindentation shows a prominent thickness-dependence, while Young's modulus of h-FeTe remains unchanged. The continuum mechanics model has been introduced to explain the observed mechanical behaviors of FeTe nanosheets. These results could shed light on the potential applications of FeTe in nanoelectronics and nanomechanics.

**RESULTS AND DISCUSSION**

According to iron-tellurium phase diagram [28], iron and tellurium can form many different structures, of which FeTe crystal can exhibit various polymorphs, such as tetragonal and hexagonal phases, suggesting its phase-dependent and diverse mechanical properties. Figure 1a demonstrates the lattice structure of layered *t*-FeTe (space group *P*4/*nmm*), where Fe layer and Te double slabs are arranged alternately in *c* direction, which provides its antiferromagnetic nature below ~70 K [29]. Our high-resolution transmission electron microscopy (HRTEM) results clearly proved the detailed atomic structures of the crystalline *t*-FeTe along [001] zone axis, as shown in Figure 1b.

The measured distances between in-plane Fe-Fe and Te-Te planes are the exactly same as 2.6 Å, in well consistent with tetragonal symmetry ($a = b$, and $\gamma = 90°$) and theoretical results [30]. The captured selected area electron diffraction (SAED) pattern displayed only one set of orthogonal diffraction spots in Figure 1c, indicating the high-quality and clean surface of t-FeTe crystal as well. The representative Raman spectrum of the as-grown t-FeTe in Figure 1d exhibited two distinct peaks at 118 and 137 cm$^{-1}$, corresponding to its $E_g$ and $A_{1g}$ phonon modes, which are in good accordance with theoretical calculations [25]. Furthermore, the spatial distributions of Fe and Te atoms were characterized by energy-dispersive X-ray spectroscopy, and the mapping results were quite homogenous throughout the entire crystal in Supplementary Figure S1. The atomic ratio Fe:Te was measured as around 1:1, in terms of its chemical stoichiometry. The tetragonal lattice was further confirmed by X-ray powder diffraction (XRD) as shown in Figure S1d.

Alternatively, *h*-FeTe (space group *P*6$_3$/*mmc*) is with non-layered ferromagnetic structure in Figure 1e, and the long-range magnetic order is aligned with *c* axis in *h*-FeTe (30 nm thick) under 220 K [29]. In stark contrast to *t*-FeTe, the octahedral vacancies in *h*-FeTe appear due to AB-stacked planes of Te atoms and are occupied by the close-packed Fe atoms, which hence leads to the non-layered nature of *h*-FeTe. It is obvious that HRTEM results revealed the in-plane six-fold symmetry ($a = b$, and $\gamma = 120°$) along [001] zone axis, as illustrated in Figure 1f. The high degree of crystallinity of our *h*-FeTe specimen was further verified by the corresponding SAED pattern, where only one set of hexagonal spots can be observed (Figure 1g). Meanwhile, two apparent Raman scattering peaks at 117 and 137 cm$^{-1}$ can be attributed to $E_g$ and $A_{1g}$ vibration modes of *h*-FeTe, in well agreement with the previous report [25].

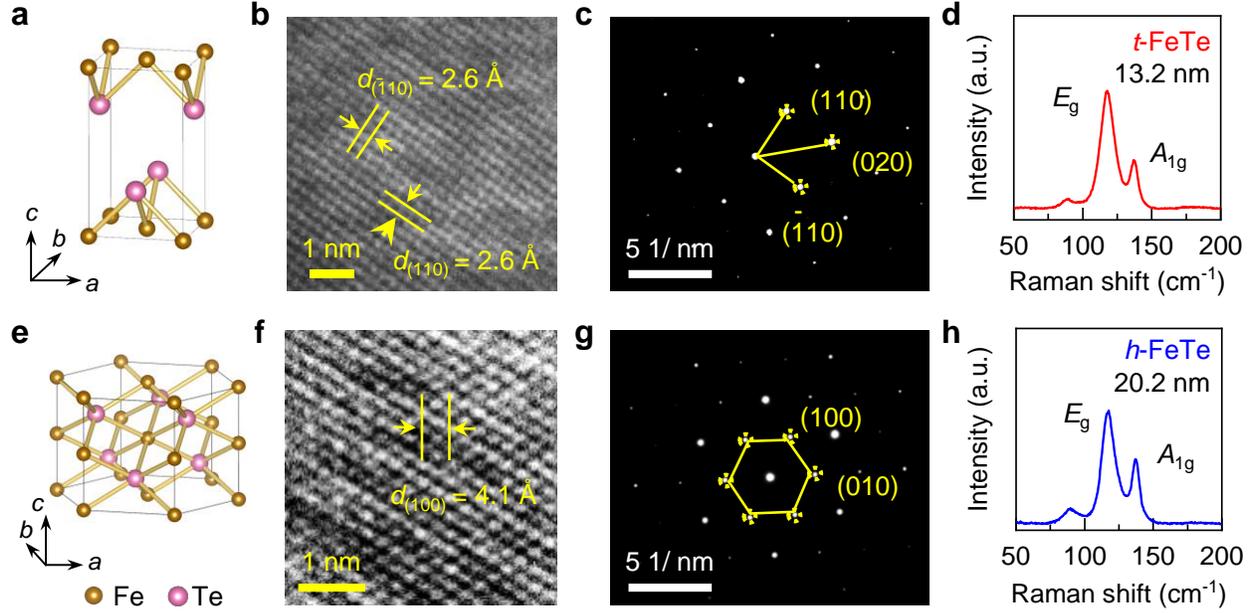

**Figure 1**. Atomic structures and Raman spectrum of FeTe nanosheets with tetragonal and hexagonal polymorphs. (**a**) Schematic diagram of crystal lattice of layered antiferromagnetic *t*-FeTe. The golden and pink balls indicate Fe and Te atoms, respectively. (**b**) HRTEM image of layered *t*-FeTe with atomic resolution. (**c**) SAED pattern of *t*-FeTe with identified (110) and (020) planes. (**d**) The representative Raman spectrum of *t*-FeTe (13.2 nm thick). (**e**) Lattice structure of non-layered ferromagnetic *h*-FeTe. (**f**) Atomic-resolved HRTEM result of *h*-FeTe. (**g**) SAED pattern of *h*-FeTe with the marked (100) and (010) planes. (**h**) Raman spectrum of *h*-FeTe. The thickness measured by AFM was 20.2 nm.

We then turn to examine the phase-modulated elastic properties of *t*-FeTe and *h*-FeTe using AFM-based nanoindentation method. As shown in Figure 2a, *t*-FeTe flakes suspended on Si substrate were realized by a dry-transfer technique, in order to exclude any potential contaminations. There was an array of holes patterned on Si surface, and their diameter and depth approximated to 250 nm and 1.5 μm, respectively. We first measured the elasticity of monolayer graphene as a reference sample, and the extracted result of 329 ± 21 N/m was quantitatively consistent with literature data [18], suggesting the great accuracy of our indentation system (see more details in Supplementary Figure S2). The thickness of *t*-FeTe was measured as 13.2 nm by AFM, and the clean surface was free of any apparent wrinkles or cracks as shown in Figure 2b. Before nanoindentation test, the FeTe flakes were cyclically scanned until the thermal drift was negligible. The cyclic load-indentation depth ($F$-$\delta$) curves were acquired with the different maximum loads ranging from 425 to 837 nN, and their consistency and repeatability confirmed a quite stable contact between $t$-FeTe and Si substrate. To derive the Young's modulus and pretension, the deformation at the center of $t$-FeTe nanosheets can be reasonably considered as an elasticity using continuum mechanical model. Importantly, the relationship between the applied force $F$ and indentation depth $\delta$ can be described as $F = \left[\frac{4\pi E}{3(1-v^2)}\left(\frac{t^3}{r^2}\right)\right]\delta + (\pi T)\delta + \left(\frac{Eq^3 t}{r^2}\right)\delta^3$ [31], where $t$ and $r$ are the thickness and radius of Si hole; $E$ and $T$ denote elastic modulus and pretension of $t$-FeTe flake, respectively, and $q$ is a dimensionless constant determined by Poisson's ratio ($v$) of the flake, obeying $q = 1.05 - 0.15v - 0.16v^2$. Taking $v$~0.21 for FeTe based on first-principle calculations [32], the fitted curve under 837 nN in Figure 2c reveals that the Young's modules and pretension were 295.3 GPa and 1.2 N/m for $t$-FeTe, respectively. Meanwhile, the inset in Figure 2c shows the typical load-indentation depth curve in logarithmic coordinates, suggesting that the indentation data of FeTe nanosheets can be well described using above equation. We further plotted the histogram of all 2D elastic modulus results obtained from a given $t$-FeTe flake, which perfectly follow Gaussian distribution with the center at 3839.7 N/m, as shown in Figure 2d (more optical images and AFM results shown in supplementary Figure S3 and S4). Notably, although the nanoindentation experiments were conducted in ambient condition, the repeatable results and plenty of control measurements demonstrated that FeTe flakes were stable enough during our testing period (more details in Supplementary Figure S5)

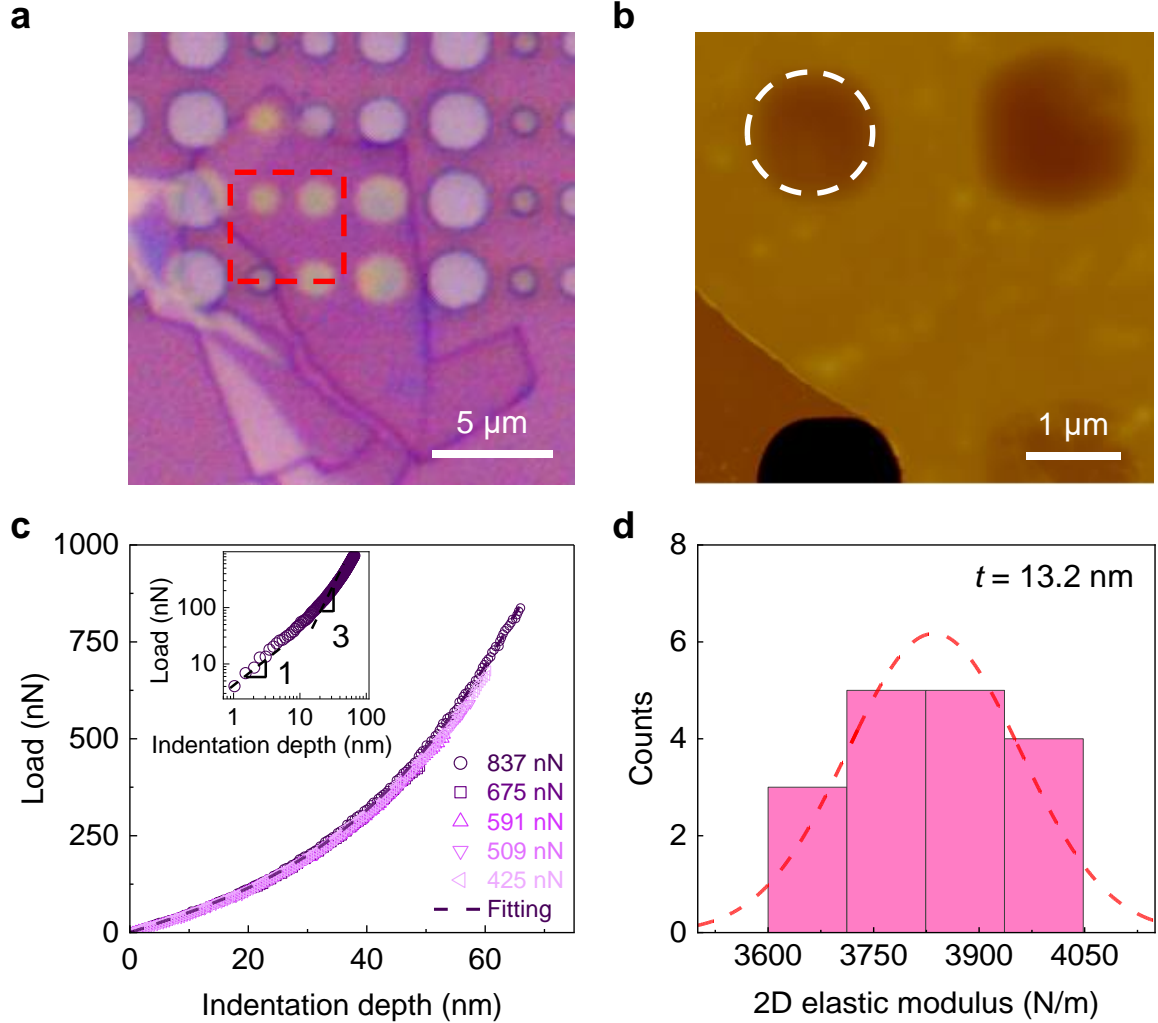

**Figure 2.** AFM nanoindentation test for elastic properties of 2D *t*-FeTe. (**a**) Optical image of *t*-FeTe flake suspended on Si/SiO$_2$ substrate with the pre-patterned array of circular holes. Red dashed square indicates the area for AFM measurement. (**b**) Topographical result of the same *t*-FeTe flake in (**a**) by AFM. White dashed circle means the zone for indentation measurement. (**c**) Five consecutive load-displacement curves of *t*-FeTe under different forces. The curve acquired under 837 nN was well fitted as marked by black dashed line. The inset shows the representative load-indentation depth curve in logarithmic coordinates. The dashed lines indicate that *F* increased linearly with *δ* first (the slope was fitted as ~1), and then *F*-*δ* displayed a cubic relationship at larger *δ* (the slope reached 3). (**d**) Histogram of the measured two-dimensional elastic modulus ($E^{2D}$) for a given *t*-FeTe (13.2 nm thick). The red dashed line represents the fitting result based on Gaussian distribution.

In comparison with *t*-FeTe, we further characterized elastic properties of the suspended 2D *h*-FeTe flakes as shown in Figure 3a. AFM scanning confirmed its thickness as 17.1 nm, which was further used to precisely position the tip at the center of circular holes in Figure 3b. Similarly, the various load-displacement tests presented the significant consistency even till the maximum load approached 500 nN, suggesting high quality of the prepared sample and structural stability of 2D *h*-FeTe. Following the same model as above, elastic modulus and pretension were extracted as 101.9 GPa and 0.5 N/m based on the linear and third-power fittings, respectively, as shown in the inset of Figure 3c. The histogram of 2D elastic modulus of *h*-FeTe obtained from different holes and loading parameters obviously suggested the elastic modulus $E^{2D}$ as 1802.9 N/m, as displayed in Figure 3d. It is noticed that *t*-FeTe and *h*-FeTe flakes showed the remarkably different mechanical behaviors, including the elastic modulus and pretension. Obviously, the Young's modulus of *t*-FeTe is dramatically 2.9 times larger than *h*-FeTe, which therefore enable us to conveniently tune the elastic properties of FeTe crystals by phase modulation.

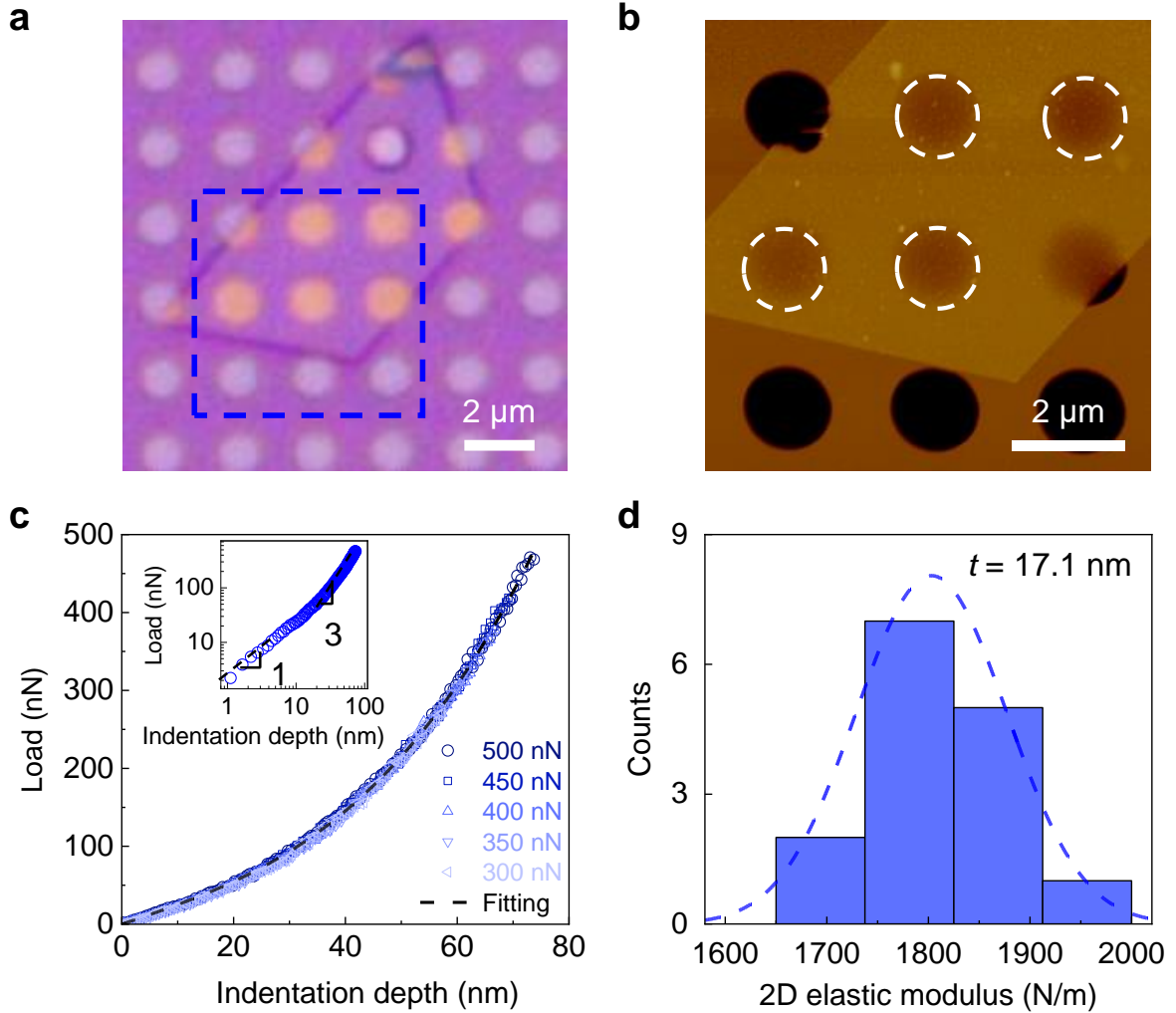

**Figure 3.** AFM nanoindentation for elastic properties of 2D *h*-FeTe. (**a**) Optical image of *h*-FeTe nanosheet suspended on Si/SiO$_2$ substrate with pre-patterned array of circular holes. (**b**) The typical AFM image of *h*-FeTe, taken from the blue dashed square in (a). (**c**) Load-indentation depth curves of *h*-FeTe under the increasing forces from 300 to 500 nN. The black dished line indicates the fitting result under 500 nN. The inset in logarithmic coordinate displays their linear and third-power relationships. (**d**) Histogram of the measured in-plane elastic modulus for a given 2D *h*-FeTe flake (17.1 nm thick). The blue dashed line means the fitting result based on Gaussian distribution.

Next, we systematically measured Young's modulus and pretension of 2D *t*-FeTe and *h*-FeTe with various thickness to emphasize its phase-dependent elastic properties. As shown in Figure 4a, Young's modulus of 2D *t*-FeTe nanosheets decreases monotonically with the increased thickness till approaching the bulk result, from the maximum 290.9±9.2 GPa for 13.2 nm thick to 103.6±11.0 GPa for 22.0 nm or thicker samples [32]. This thickness-dependence of 2D *t*-FeTe is resulted from the potential interlayer sliding due to the weak van der Waals coupling, which occurred in layered b-P as well [33-35]. As the layer number becomes larger, each layer may choose a specific in-plane sliding path to minimize the total energy, which contribute to the lower elastic modulus in thicker *t*-FeTe. Obviously, the layer-dependence of elastic modulus in *t*-FeTe is significantly stronger than other known 2D materials in Figure 4a. In contrast, Young's modulus of 2D *h*-FeTe nanosheets is almost independent of thickness, which is quite different from 2D layered *t*-FeTe. Based on the non-layered nature of 2D *h*-FeTe, the external load can't break its chemical bonds, so the measured modulus remained unchanged even for thin flakes in our case. Furthermore, Figure 4b presents the pretension of *t*-FeTe and *h*-FeTe nanosheets with different thicknesses. The extracted pretension ranges from 0.11 to 1.1 N/m and 0.06 to 0.75 N/m for *t*-FeTe and *h*-FeTe, respectively, which maybe induced during transfer process [36].

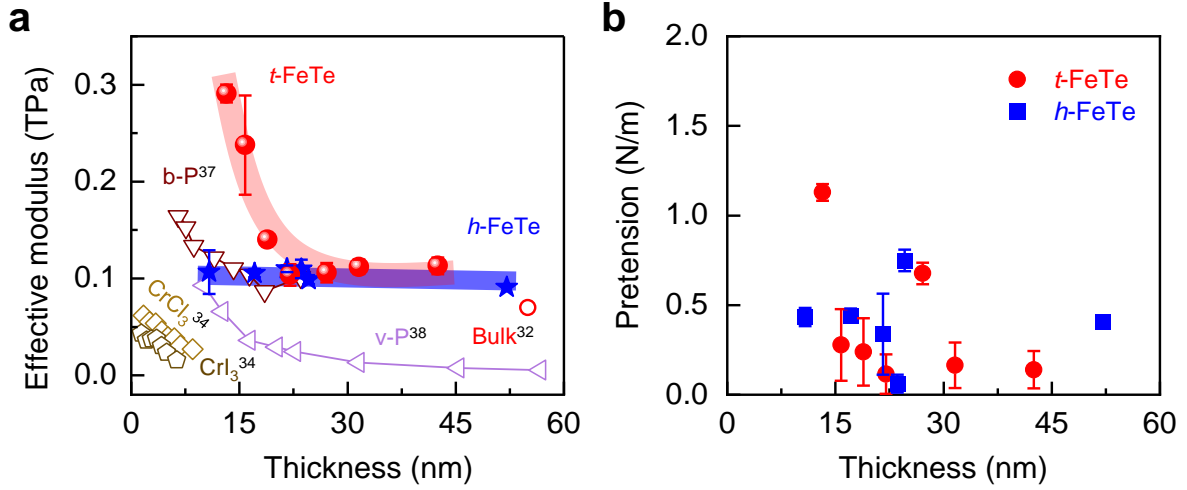

**Figure 4.** Thickness-dependence of Young's modulus and pretension of 2D *t*-FeTe and *h*-FeTe nanosheets and comparison with other 2D materials. (**a**) Comparison of effective modulus of 2D materials, including *t*-FeTe, *h*-FeTe, b-P [37], violet phosphorus (v-P) [38], $CrI_3$ [34], and $CrCl_3$ [34]. It is obvious that effective modulus of 2D *t*-FeTe and *h*-FeTe show the significantly different layer-dependence. The references are labeled with black numbers. The elastic modulus of bulk FeTe (red circle) is calculated based on polycrystal [32]. (b) The pretension in *t*-FeTe and *h*-FeTe nanosheets with different thicknesses.

## CONCLUSIONS

In summary, we have measured the elastic properties of suspended two-dimensional magnetic FeTe nanosheets, and computed the elastic modulus and pretension with continuum models. With the increase in thickness, the elastic modulus of 2D antiferromagnetic *t*-FeTe dramatically decreases due to interlayer slippage. In comparison, elastic modulus of the ferromagnetic *h*-FeTe essentially unchanged with the increase in thickness due to its non-layered structure. Meanwhile, when *t*-FeTe thickness reaches 22 nm, its elastic modulus approaches to that of bulk FeTe. We believe that these results could boost the potential applications of 2D magnetic FeTe materials in flexible devices and nano-spintronics.

## EXPERIMENTAL METHODS

**Preparation and characterization of the suspended 2D FeTe flakes**

The 2D FeTe flakes with tetragonal and hexagonal phases were rationally grown on mica by the optimized chemical vapor deposition (CVD) as reported elsewhere [25]. To facilitate AFM and TEM measurements, a modified dry-transfer method based on polymer template [25] was developed in this work. First, non-polar polystyrene (PS) particles (~13 g) were dissolved in toluene (~100 mL). Then, 2D FeTe flakes on mica surface were coated by a thin layer of PS by spin-coating technique (speed was 2000 rpm), followed by thermal baking at 70 °C for ~15 min. Second, we scratched the edges of PS/FeTe/mica for the fast separation by a sharp blade, and then immersed PS/FeTe/ mica in ultrapure water. After a while, the PS with sample was floated on the surface of water. Afterwards, PS/FeTe was picked up gently by a sharp tweezer, rinsed by DI water for several times, and subsequently transferred to fully cover the circular holes on a Si wafer ( 300 nm oxidization layer). The holey Si wafer was prepared by UV lithography and dry etching by $CF_4$. The typical dimensons of those holes were ~1.0 μm diameter, 0.25 μm depth, and 2.5 μm interval. A lacey cabron grid was used to pick up FeTe flake for TEM measurements. Finally, PS layer was easily removed with toluene, leaving the clean 2D FeTe on surface.

The high quality of CVD grown crystalline FeTe was preliminarily characterized by optical microscopy (Olympus BX53) and Raman spectroscopy (Horiba iHR550 spectrometer, and the excitation laser was 632.8 nm). Flake thickess was accurately determined by AFM (Dimension Icon, Bruker). The lattice strucutres of FeTe flakes were measured by XRD (X-Pert3 Powder with $K_α$ radiation of Cu) and TEM (JEM-2010F, acceleration voltage 200 kV).

**AFM nanoindentation**

For AFM nanoindentation tests, the load localized around the tip was applied to acquire the load-indentation depth curve using PF-QNM mode in AFM (Dimension Icon, Bruker). In general, spring constant *k* of silicon cantilever of AFM tip can be precisely calibrated by the well-known Sader method, following a harmonic oscillation model (more details are given in Figure S6) [22]. In our case, the represnetative spring constant and frequency were 40 N/m and 300 kHz, respectively, as shown in Figure S7. Each FeTe flake was scanned for several times till thermal drift effect was negligible before recording the load-indentation depth data. AFM morphology did not show any evident slippage of FeTe from Si substrate. For indentation test, $z$-piezo displacement speed was tuned at a rate of 300 nm/s in load-indentation depth curve. The indentation depth $\delta$ was defined as the central displacement of the suspended FeTe flake. The applied load $F$ can be calculated by Hooke's law, $F = kx$, where $x$ was the deformation of probe cantilever and directly given by AFM system. Thus, indentation depth $\delta$ of the suspended sample can be derived as $\delta = z - x$, where $z$ is the moving distance between the tip and sample.

**Analysis of load-displacement curves**

In equation $F = \left[\frac{4\pi E}{3(1-v^2)}\left(\frac{t^3}{r^2}\right)\right]\delta + (\pi T)\delta + \left(\frac{Eq^3 t}{r^2}\right)\delta^3$ [39], it included three terms to calculate the pretension and Young's modulus of 2D FeTe nanosheets: The first term corresponds to the mechanical behavior of a flake with a certain bending rigidity in equation, the second term means the pretension of membrane, and the third one corresponds to the stiffness of FeTe flakes under large deformation. When the tip moved downward, the cantilever bent and $z$-piezo displacement was recorded during the indentation experiments. In general, before the force-indentation test, the cantilever bending was calibrated by measuring the force-displacement curve on $SiO_2$/ Si surface. This step was repeated until the trace and retrace curves were exactly overlapped.

In order to obtain the intrinsic elastic modulus of 2D FeTe, it is of importance to correct load-indentation depth curve. In principle, as AFM tip moving very close to sample surface, it can rapidly snap to FeTe due to van der Waals interaction, which inevitably affect the load-indentation depth data. In this scenario, a negative force is principally contributed to the real force-deflection data. Therefore, a critical point was defined by extrapolating the zero-force line before the snapping behavior, where the force and displacement are both zero. A detailed example can be found in Supplementary Figure S8.

**ASSOCIATED CONTENT**

**Supporting Information**

The Supporting Information is available in the online version of this article.

Optical images and EDS mapping results of FeTe nanosheets, XRD results of *t*-FeTe nanosheets, optical images of *t*-FeTe and *h*-FeTe with different thickness, AFM images of *t*-FeTe and *h*-FeTe flakes before and after nanoindentation, SEM image of AFM tips, calibration of spring constant $k$ of the AFM tip, mechanical properties of graphene monolayers, analysis of force-indentation curves, and sample stability test under ambient condition.

**Author Contributions**

Y.C. and Y.Y. conceived this project and designed the experiment. Y.Y., G.D., and W.H. prepared the suspended samples and performed nanoindentation tests. M.C. and J.S. carried out the CVD growth of FeTe flakes. Z.T. and Y.G. grew the FeTe crystals. Y.C. and Y.Y. wrote the manuscript with the necessary input of all authors. All authors have given approval to the final manuscript.

**Notes**

The authors declare no competing financial interests.

**Acknowledgements**

This work was financially supported by the National Natural Science Foundation of China (grant numbers 52072032, 12090031, and 92164103) and the 173 JCJQ program (grant number 2021-JCJQ-JJ-0159). Y.G. acknowledges the support by the Double First-Class Initiative Fund of ShanghaiTech University.

*Supplementary Information*

# Phase-Modulated Elastic Properties of Two-Dimensional Magnetic FeTe: Hexagonal and Tetragonal Polymorphs

Yunfei Yu,[1] Mo Cheng,[2] Zicheng Tao,[3,4] Wuxiao Han,[1] Guoshuai Du,[1] Yanfeng Guo,[3,4] Jianping Shi,[2] Yabin Chen[1,5,6,*]

[1]*School of Aerospace Engineering, Beijing Institute of Technology, Beijing 100081, P.R. China*

[2]*The Institute for Advanced Studies, Wuhan University, Wuhan 430072, P.R. China*

[3]*School of Physical Science and Technology, ShanghaiTech University, Shanghai 200031, P.R. China*

[4]*ShanghaiTech Laboratory for Topological Physics, Shanghai 201210, P.R. China*

[5]*Advanced Research Institute of Multidisciplinary Sciences, Beijing Institute of Technology, Beijing 100081, P.R. China*

[6]*BIT Chongqing Institute of Microelectronics and Microsystems, Chongqing, 400030, P.R. China*

[*]Corresponding author: Y.C. (chyb0422@bit.edu.cn)

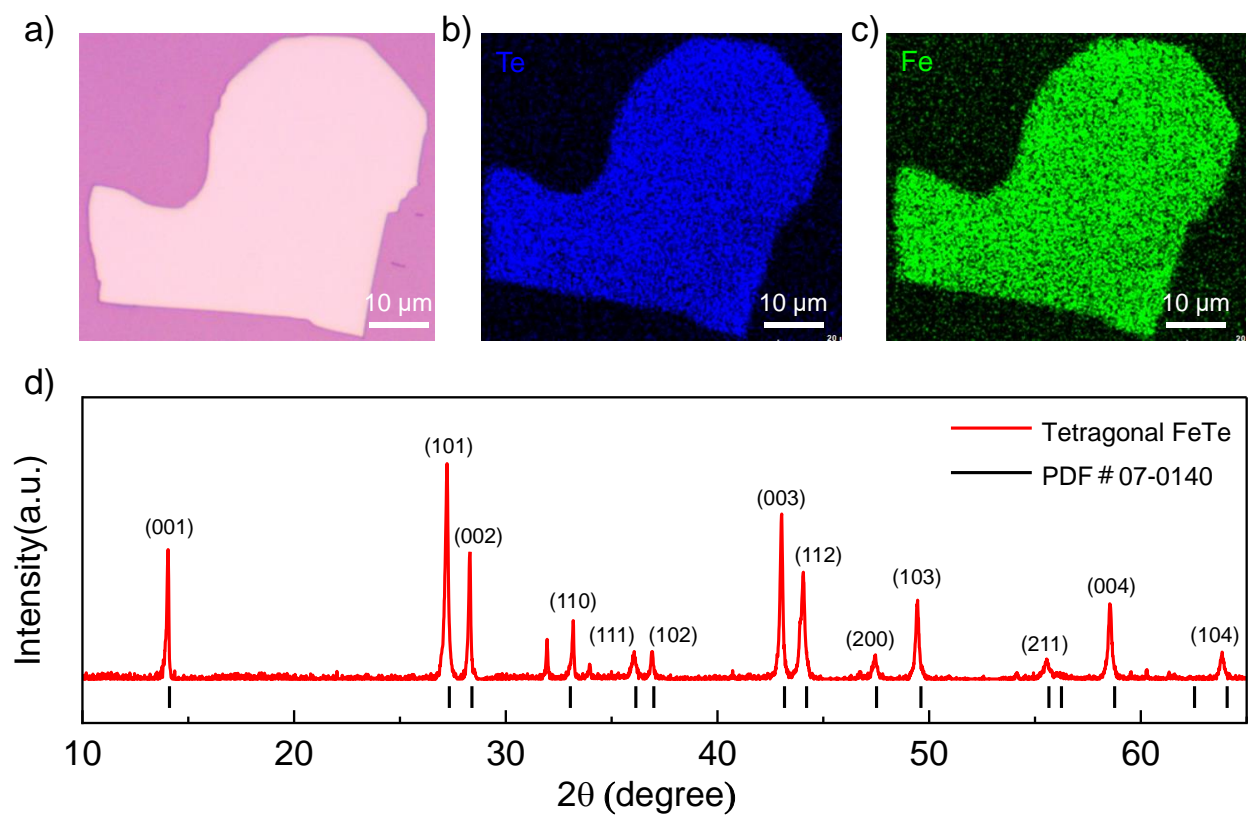

**Figure S1.** Structural characterizations of 2D *t*-FeTe. (**a**) Optical image of *t*-FeTe nanosheets. (**b-c**) Corresponding EDS mapping results of Te (**b**) and Fe (**c**) elements for *t*-FeTe nanosheets, respectively. The atomic ratio of Fe and Te elements approaches 1:1. (**d**) XRD pattern of *t*-FeTe, well consistent with PDF results. The sharp peaks indicate high quality of FeTe crystal.

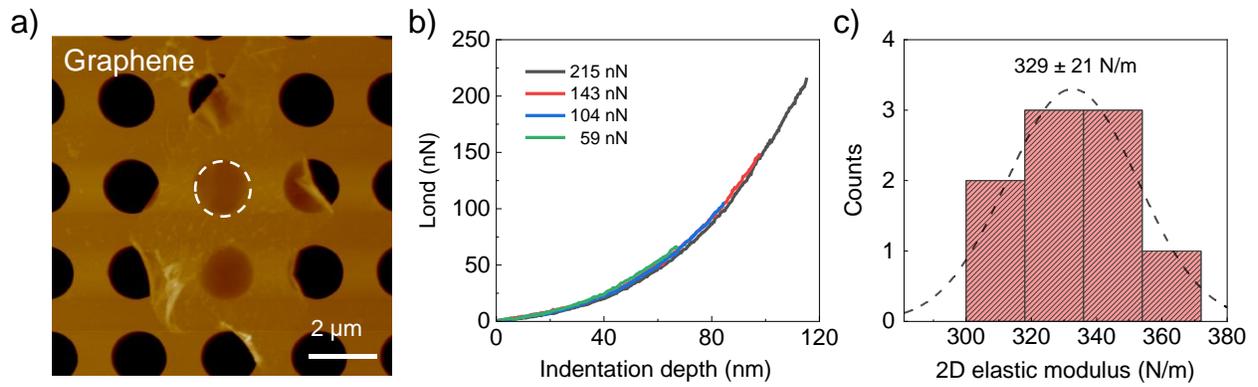

**Figure S2.** Monolayer graphene used as standard sample to calibrate the indentation setup (**a**) Representative AFM image of a graphene flake suspended on the fabricated holes of Si substrate. (**b**) Force-indentation curves of monolayer graphene under different loads. All curves are consistent with each other. (**c**) Histogram of the measured elastic modulus for monolayer graphene. Dotted line represents the Gaussian fitting, and the central result is 329 ± 21 N/m.

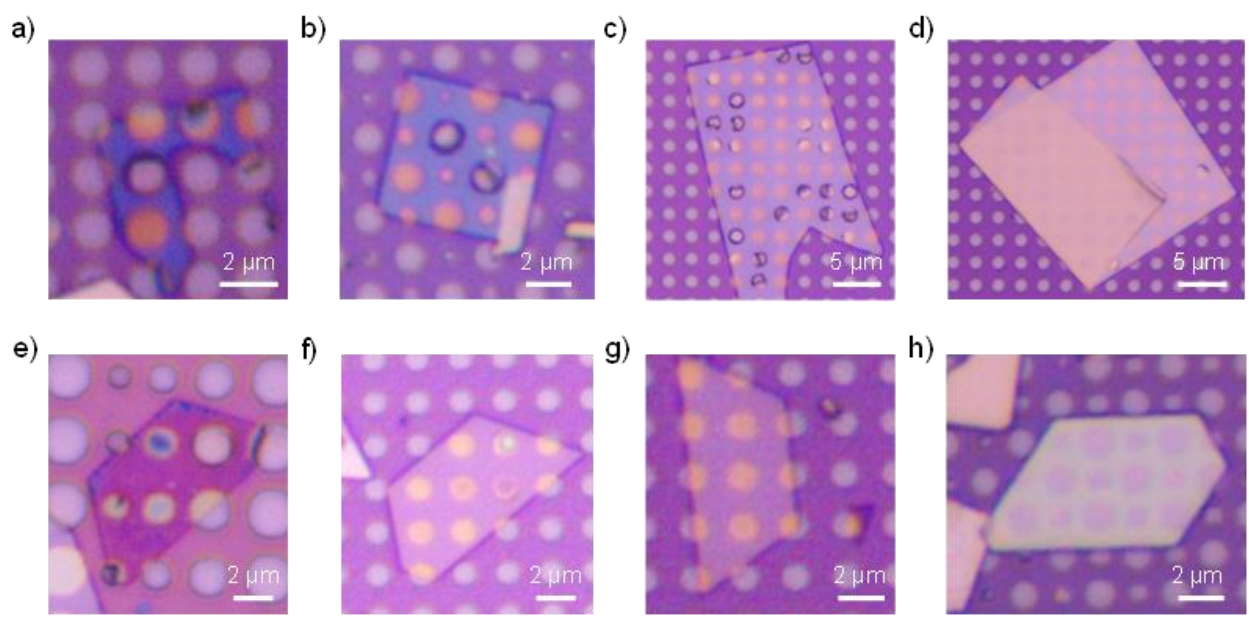

**Figure S3.** Optical images of 2D *t*-FeTe and *h*-FeTe with different thickness. (**a-d**) Optical images of *t*-FeTe flakes with different thickness of 18.9 (**a**), 22.0 (**b**), 31.6 (**c**), and 42.5 nm (**d**), respectively. (**e-h**) Optical images of 2D *h*-FeTe with various thickness of 10.8 (**e**), 21.6 (**f**), 23.6 (**g**), 52.1 nm (**h**), respectively. The flake thickness was determined by AFM.

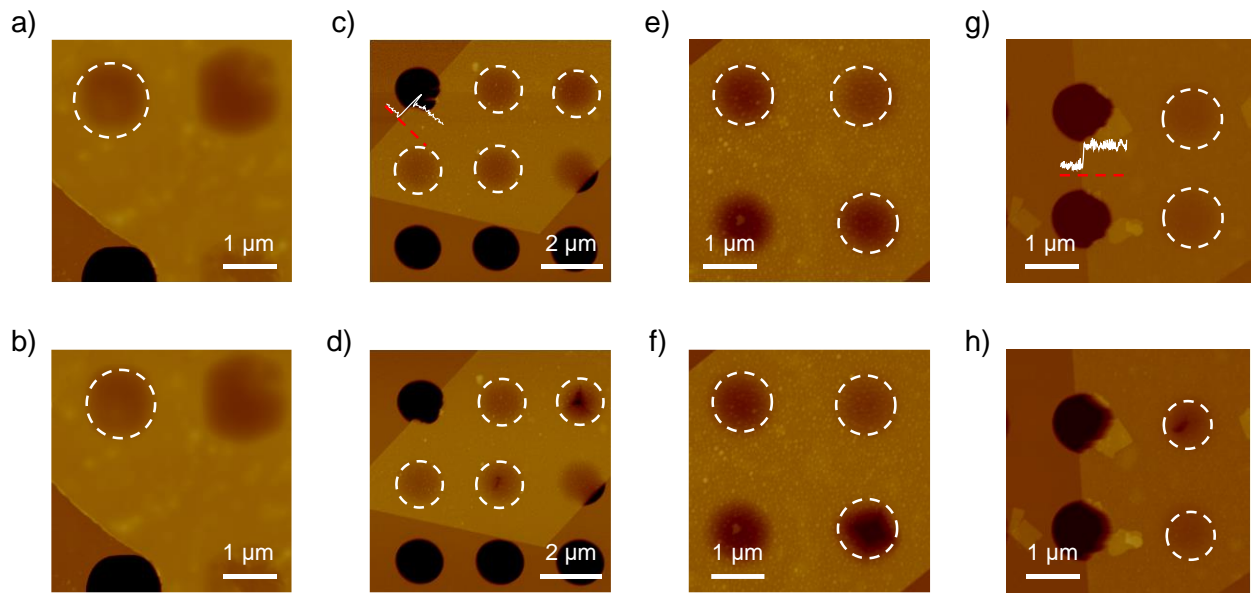

**Figure S4.** AFM images of *t*-FeTe and *h*-FeTe with different thickness before and after nanoindentation tests. (**a, b**) AFM images of *t*-FeTe (13.2 nm) before and after experiment, respectively. (**c-h**) AFM images of *h*-FeTe with 17.1 nm (**c, d**), 21.6 nm (**e, f**), 23.6 nm (**g, h**) before and after indentation, respectively. The white dotted circles indicate the tested samples.

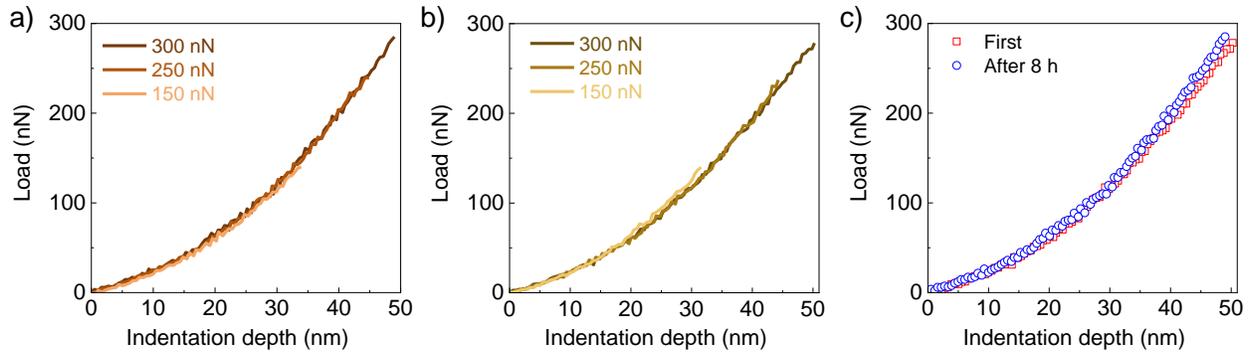

**Figure S5.** Sample stability test under ambient condition. (**a**) The first nanoindentation results under different forces in ambient condition. (**b**) The second nanoindentation results of the same sample in (**a**) after 8 hours in ambient condition, using the same experimental parameters. (**c**) Two repeated load-indentation depth curves, suggesting that FeTe flakes are quite stable during the test period. Red and blue data represent the first and second load-indentation depth ($F$-$\delta$) curves, respectively.

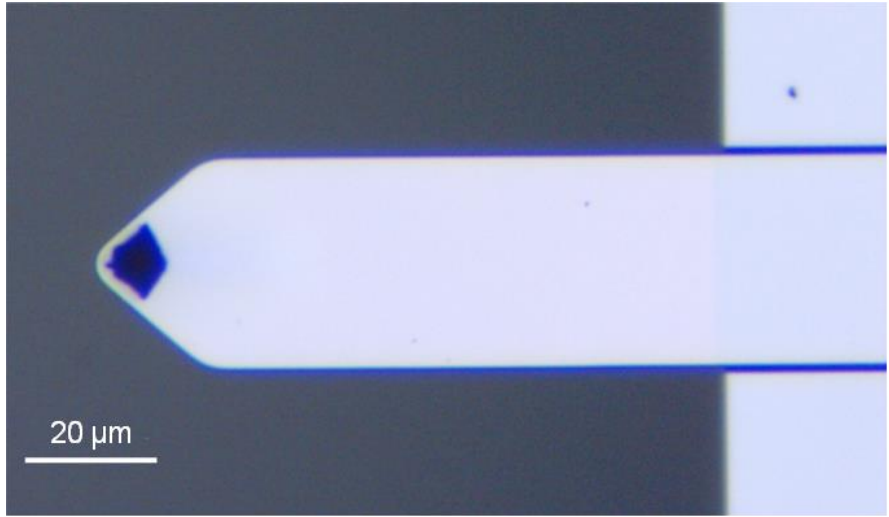

**Figure S6.** Optical image of the AFM probe.

Spring constant $k$ of the AFM probe is defined as $k = \frac{F}{x}$, and can be determined by Sader method[1], which states:

$$k = 0.1906 \rho b^2 L Q_f \omega_f \Gamma_i^f(\omega_f)$$

where $L$ and $b$ are the length and width of cantilever which are measured directly by optical microscope; $\rho$ is the density of the fluid (air, 100 kPa, 298 K); $\omega_f$ and $Q_f$ are the radial resonant frequency and quality factor of the fundamental resonance peak, which are determined in the AFM system following simple harmonic oscillator; $\Gamma_i^f(\omega_f)$ is the imaginary part of the hydrodynamic function. For a simple calibration of spring constant, the online calibration is very helpful (https://ampc.ms.unimelb.edu.au/afm/webapp.html), as suggested by the AFM manufacture.

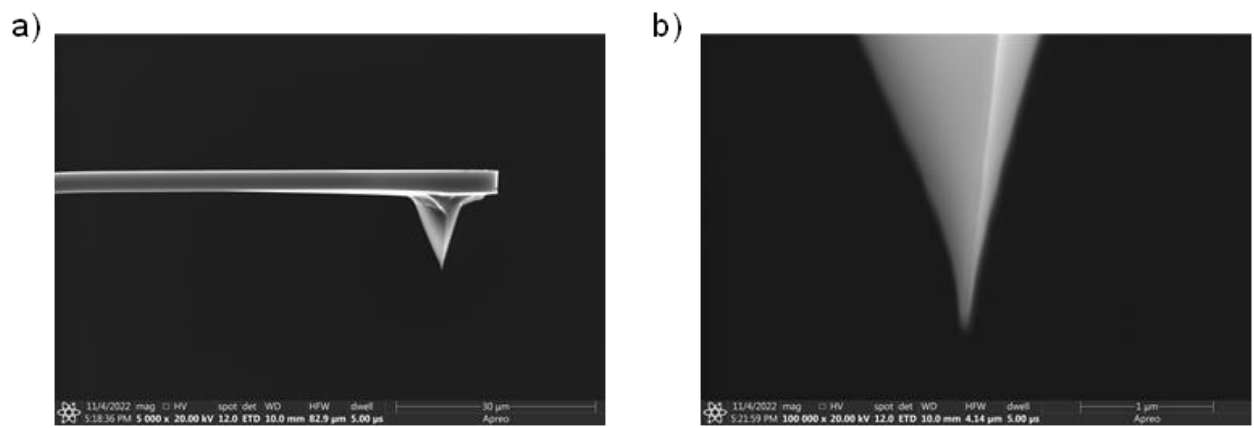

**Figure S7.** SEM images of AFM tips with low (**a**) and high (**b**) magnifications.

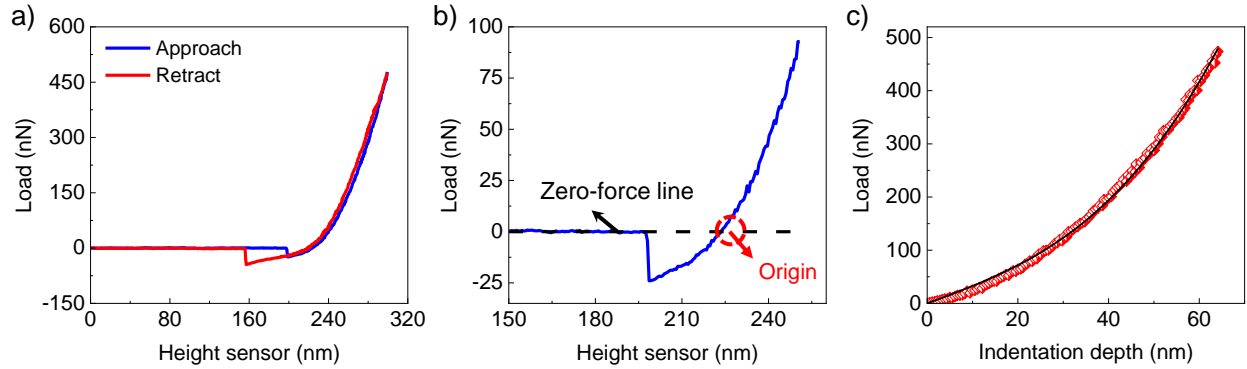

**Figure S8.** The detailed analysis of force-indentation curves. (**a**) The representative approaching and retracting force curves on *t*-FeTe nanosheet with thickness of 15.8 nm. (**b**) A typical curve of load versus scanning piezo displacement for *t*-FeTe flake. We extrapolate the black dashed line (zero-force line) until it crosses the blue curve, and confirm this intersection point (the red dotted circle) as the origin where both load and indentation depth are zero. (**c**) The least-square fitting of the experimental indentation data was carried out by using the equation in main text.